\documentclass[iop,numberedappendix, twocolappendix]{emulateapj}

\usepackage{epsfig,wrapfig,color,sidecap}
\usepackage{index}
\usepackage{subfigure}
\usepackage{enumerate}
\usepackage{graphicx}
\usepackage{times}
\usepackage{amssymb}
\usepackage{color}
\usepackage{amssymb,amsmath}
\usepackage{colordvi}
\usepackage{times}

\usepackage{colordvi}

\RequirePackage[colorlinks=true
,urlcolor=blue
,anchorcolor=blue
,citecolor=blue
,filecolor=blue
,linkcolor=blue
,menucolor=blue
,pagecolor=blue
,linktocpage=true
,pdfproducer=medialab
]{hyperref}

\def\thetamax{\theta_{\textrm{max}}}
\def\etacar{$\eta$-Carinae}

\begin{document}


\title{On Continuum-driven Winds from Rotating Stars}

\shortauthors{Shacham \& Shaviv}

\author{Tomer Shacham \& Nir J. Shaviv}
\affil{Racah Institute of Physics, The Hebrew University of Jerusalem, \\ Jerusalem, 91904, Israel}

\begin{abstract}
We study the dynamics of continuum driven winds from rotating stars, and develop an approximate analytical model.  We then discuss the evolution of stellar angular momentum, and show that just above the Eddington limit, the winds are sufficiently concentrated towards the poles to spin up the star. A twin-lobe structure of the ejected nebula is seen to be a generic consequence of critical rotation. We find that if the pressure in such stars is sufficiently dominated by radiation, an equatorial ejection of mass will occur during eruptions. These results are then applied to $\eta$-Carinae. We show that if it began its life with a high enough angular momentum, the present day wind could have driven the star towards critical rotation, if it is the dominant mode of mass loss. 
We find that the shape and size of the Homunculus nebula, as given by our model, agree with recent observations. Moreover, the contraction expected due to the sudden increase in luminosity at the onset of the Great Eruption explains the equatorial ``skirt" as well. 
\end{abstract}


\section*{Introduction}

\etacar\ epitomizes the class of Luminous Blue Variable (LBV) stars. During the 1840's Great Eruption, 
it shed several solar masses while radiating at several times the Eddington luminosity \citep{SmithFrew}. The Homunculus  nebula which formed exhibits a clear bipolar structure with an equatorial ``skirt" \citep[e.g.,][]{DavidsonReview}.

Several ideas have been suggested to explain the broken spherical symmetry. One type of mechanisms relates the ejected nebula to the rotation of \etacar. A second type links the 
nebula to an interaction with the companion star.

\cite{Frank1995} described the Homunculus as the result of a fast nearly spherical wind interacting with a slow, previously accelerated, equatorial wind from a rotating star. \cite{Frank1998} describes an opposite model, where a fast aspherical wind interacts with a spherical one. 

\cite{Owocki1998} and \cite{Maeder1999} pointed out that von Zeipel's theorem \citep{vonZeipel1924} has a significant effect on the geometry of the wind. In particular, mass loss from a fast rotator is expected to be predominantly concentrated towards the poles, where the effective gravity and radiative field are large. This naturally explains the bipolar structure without requiring a two winds interaction, as previously suggested. 

\cite{Maeder2001} applied the standard line driven winds model \citep{CAK} to explain the form of the nebula. They have shown that if the polar temperature is $\sim 2.5\times10^4\,$K, the line driving bi-stability jump would give rise to a large increase in the equatorial mass loss, thus offering an explanation to the origin of the ``skirt".  

An alternative model for the overall structure of the Homunculus was developed by  \cite{Smith2007}. The nebula is described as explosively ejected material from a critically rotating surface. An assumption that the initial velocity is proportional to the effective gravity gives the bipolar structure. The skirt is obtained within some range of parameters, for which the orbits of matter originating at opposite low latitudes pass through the equatorial plane and collide.   

Recently, \cite{Soker2004,Soker2005} advocated the idea that the twin lobe structure is due to an interaction with the companion star. In particular, Soker argued that it is hard to reconcile the large bipolar mass loss of the Great Eruption with {\em standard} models of stellar winds, for two reasons. First, one cannot obtain a high ratio between the wind momentum ($\dot{M}_w v_\infty$) and the photon momentum ($L/c$), as observed. Second, in order to explain the bipolar structure, the envelope is assumed to be near critical rotation. 
However, this is inconsistent with large mass loss, as it would slow down the envelope considerably. 
Soker also presented a model in which a slow wind from $\eta$-Carinae is accreted onto the secondary, accelerating twin jets which form the Homunculus. As we shall see below, a model of super-Eddington (SED) continuum driven winds from a rotating star circumvents both problems. 

 \cite{ShavivEta} has shown that the SED state can be explained by a reduced effective opacity resulting from a ``porous" atmosphere. In subsequent work, it was shown that such atmospheres give rise to optically thick winds with a relatively simple mass-loss luminosity relation, which also explains the behavior of classical novae \citep{ShavivNovae,Owocki2004}. The bipolar structure of the Homunculus is a clear incentive to consider  that its origin is a SED continuum driven wind from a rotating star. 

This note is organized as follows. In \S\ref{Part1}, we develop the theory behind continuum driven winds from rotating stars and the associated evolution of angular momentum. In \S\ref{Part2}, this theory is then applied to the case of \etacar, explaining its high spin, and the formation of the Homunculus nebula and equatorial skirt.

As a side note, we study in \S\ref{sec:spin} generic 
interactions between the binary components, and show that spin-orbit coupling could not have been responsible for the high spin.  In \S\ref{sec:AngMomDif}, we show that during the first few years of the Great Eruption, \etacar\ could not have been rigidly rotating.

\section{Part I - Theory}
\label{Part1}

We begin in \S\ref{sec:WindsTheory} by developing an approximate analytical model 
of continuum driven winds from rotating stars. In \S\ref{sec:Spinup}, we show that at low mass loss rates, such winds are capable of spinning up the star, even though angular momentum is lost to the wind. In \S\ref{sec:SkirtTheory}, we study the consequences of eruptions to~the~evolution~of~the~stellar angular momentum.

\subsection{Continuum-driven Winds from Rotating Stars}
\label{sec:WindsTheory}

A complete analysis of the dynamics of LBV eruptions is quite complicated. In general, one should solve the Navier-Stokes equations for the wind, coupled to the radiative transfer. Furthermore, as we argue that the star is rotating near breakup, its oblique shape must be taken into account.  Such an analysis is amenable only by a numerical treatment. Since we wish to understand the underlying physics, this path will not be taken here.

In order to encapsulate the relevant phenomena within a simple analytical model, we make the following simplifying assumptions. 
\begin{enumerate}[I.]
\item The star is spherical.
\item There is no latitudinal motion ($v_{\theta}\ll v_r)$.
\end{enumerate}
Under these approximations, the model is integrable. 

We first note that the typical wind velocity is of the order of magnitude of the escape velocity, which is much larger than the speed of sound. This implies that the kinetic energy density is significantly higher than the internal energy of the gas; mechanical pressure and viscosity will therefore be neglected. 

We focus on two fields: 
the velocity of the wind $\bf v$ and the radiative flux $\bf F$, and begin with conservation of momentum:
\begin{equation}
\rho\left(\frac{\partial}{\partial t}+{\bf v}\cdot {\nabla}\right){\bf v}=\sum{\bf f}.
\label{eq:MomentumConservation}
\end{equation}
Here $\sum {\bf f} = \rho({\bf g}_\mathrm{grav} + {\bf g}_\mathrm{rad})$ is the sum of the different forces per unit mass, with
\begin{equation}
{\bf g}_\mathrm{grav}=-g\left(\frac{R}{r}\right)^{2}\hat{\bf r},\qquad {\bf g}_\mathrm{rad} =\frac{\kappa}{c}{\bf F},
\end{equation}
where $g\equiv GM/R^{2}$  and $ \kappa$ is the  opacity. At a steady state, the radial component of eq.~(\ref{eq:MomentumConservation}) is\footnote{Under these approximations the effective gravity is radial.}  
\begin{equation}
v_{r}\frac{\partial}{\partial r}v_{r}-g_{\textrm{eff}}=\frac{\kappa}{c}F_{r},
\end{equation}
where 
\begin{equation}
g_{\textrm{eff}}\equiv\frac{v_{\phi}^{2}}{r}-\frac{GM}{r^{2}}.
\end{equation}
Conservation of angular momentum implies that
\begin{equation}
v_{\phi}=\left(\frac{R^{2}}{r}\right)\omega \sin\theta,
\label{eq:AngularMomentumConservation}
\end{equation}
and hence 
\begin{equation}
g_{\textrm{eff}}=-g\left[\left(\frac{R}{r}\right)^{2}-\left(\frac{R}{r}\right)^{3}{\Omega}^{2}\sin^{2}\theta\right],
\end{equation}
where 
\begin{equation}
\Omega\equiv\omega\sqrt{{R^{3}}/{GM}}
\end{equation}
is the spin in units of the classical breakup limit. 

The radiation field loses energy as it accelerates the wind, lifting it over the gravitational pull. Conservation of energy then implies
\begin{equation}
{\nabla}\cdot{\bf F}=-\frac{\kappa}{c}\rho{\bf F}\cdot{\bf v},
\end{equation}
which reduces to
\begin{equation}
\frac{1}{r^{2}}\frac{\partial}{\partial r}\left(r^{2} F_{r}\right)=-\frac{\kappa}{c}\rho F_{r} v_{r},
\label{eq:flux}
\end{equation}
assuming axial symmetry and neglecting latitudinal motion. 

Conservation of mass, 
\begin{equation}
\frac{\partial \rho }{\partial t} + \nabla \cdot \left( \rho {\bf v} \right)=0,
\label{continuity}
\end{equation} 
closes the set of equations.
At a steady state with the assumed symmetries, it reduces to
\begin{equation}
\frac{\partial}{\partial r}\left(r^{2} \rho v_{r}\right)=0.
 \label{eq:Ftheta}
\end{equation} 

For convenience, we define
\begin{equation}
f\left(\theta \right)\equiv \frac{4 \pi r^{2} \rho v_{r}}{-\dot M},
\label{fDef}
\end{equation} 
where $\dot{M}$ is the mass loss rate and is therefore negative.\footnote{This choice was made to keep $f \left(\theta \right)=1$ in the case of no rotation.} Plugging this into eq.~(\ref{eq:flux})  and integrating gives
\begin{equation}
F_{r}= \frac{L}{4\pi r^{2}}  e^{{-m  f (\theta)}\left(1- {R}/{r}\right)} \chi(\theta),
\label{eq:Fr}
\end{equation} 
where the luminosity $L$ was introduced by dimensional analysis in order for the constant of radial integration $\chi(\theta)$ to be dimensionless,  and 
\begin{equation}
m\equiv\frac{-{{\dot{M} \kappa}}}{{{4 \pi c R}}}
\end{equation}
is the ``photon tiring number".\footnote{This definition is slightly different from the original one defined by \cite{OwockiGayley1997}. Here the mass loss is compared with the Eddington luminosity, rather than the actual one.}

Using von Ziepel's theorem \citep{vonZeipel1924} with the simplification of having just a radial flux, we can determine the radiative and mass fluxes at the base of the wind:
\begin{equation}
F_{r}\Big{|}_{r=R}\propto g_{\textrm{eff}}\Big{|}_{r=R}.
\end{equation}
The requirement $\int {\bf F}\cdot {\bf dS} = L $ gives
\begin{equation}
\chi\left(\theta \right)=\frac{1-{\Omega}^{2}\sin^{2}\theta}{1-\frac{2}{3}{\Omega}^{2}}.
\end{equation} 

Unlike line driven winds, the mass flux expected in a SED continuum driven wind is inherently local. This is because the  critical surface of the wind, where the gravitational and radiative forces balance, depends only on the size of the inhomogeneous structure formed by radiative hydrodynamic instabilities. Since the hydrostatic scale height, over which these instabilities operate, is much smaller than the stellar radius, mass loss is determined by local conditions. As the local atmospheric structure cannot distinguish between gravity and the centrifugal force, the mass loss will depend on the local {\em effective} gravity. 

Generalizing the local mass flux derived by \cite{ShavivNovae}, we have
\begin{equation}
{\bf \Phi} = \frac{\cal W} {c v_s}\left({\bf F} - {\bf F_{\textrm{crit}}}\right),
\end{equation}
where 
\begin{equation}
{\bf F_{\textrm{crit}}}={\bf F_{\textrm{edd}}}\left(1-\Omega^{2}\sin^{2}\theta\right).
\end{equation}

This is analogous to the local wind flux in SED accretion disks \citep{DotanDisks}. 
Here $F_{\textrm{edd}} = g c / \kappa$ is the ``Eddington flux", for which the radiative flux $\bf F$ balances gravity in the case of no rotation; $\bf F_{\textrm{crit}}$ is the equivalent flux when rotation is added. Last, ${\cal W}$ is the wind constant, which encapsulates mostly geometrical features described by \cite{ShavivNovae}, but may have a weak dependence on $\Gamma \equiv F/F_{\textrm{edd}}$. 

\cite{Owocki2004} have shown that this dependence is stronger if the inhomogeneous structure comprising the unstable atmosphere is power law dominated by smaller scales. However, as we shall see below, this merely scales the mass loss by a constant factor since $\Gamma$ is latitudinally independent. Note that a weak latitudinal dependence of the mass loss should arise from the latitudinal dependence of $v_s$ at the base of the wind. However, $v_s$ is proportional to $\sqrt{T}$ which itself is a very weak function of the optical depth. 

Neglecting the aforementioned weak dependences, the flux depends on $\Gamma$ as 
\begin{equation}
{\bf \Phi}=\rho{\bf  v} \propto {\bf F-F_{\textrm{crit}}}\propto \left( \Gamma -1+\frac{2}{3}\Omega^{2} \right )  \chi\left(\theta \right).
\label{eq:MassFlux}
\end{equation} 
Under the above approximations, a comparison with eq.~(\ref{fDef}) reveals that
\begin{equation}
 f\left(\theta \right)= \chi\left(\theta\right).
 \end{equation}

A surface integration of ${\bf \Phi}$,
\begin{equation}
-\dot{M}=\int {\bf \Phi} \cdot {\bf dS},
\label{eq:MdotPhi}
\end{equation}
gives 
\begin{equation}
m=\frac{1}{2}\mathcal{W}\frac{v_{\textrm{esc}}^{2}}{c\, v_{s}}\left(\Gamma-1+\frac{2}{3}\Omega^{2}\right)\equiv \tilde{\mathcal{W}} \left(\Gamma-1+\frac{2}{3}\Omega^{2}\right),
\end{equation}
where $\tilde{\mathcal{W}}$ is the scaled wind constant and
$v^{2}_\textrm{esc}\equiv2GM/R$. 

An explicit form of $F_{r}$ (eq.~\ref{eq:Fr}) allows a direct integration of $v_{r}$ (eq.~\ref{eq:flux}):
\begin{equation}
\frac{1}{2}v_{r}^{2}\Big|_{R}^{r}=\int_{R}^{r}dr\left(g_{\textrm{eff}}+\frac{\kappa}{c}F_{r}\right).
\end{equation}
Neglecting the velocity at the base of the wind gives
\begin{eqnarray}
\left(\frac{v_{r}}{v_\textrm{esc}}\right)^{2} &=& \left(1-\frac{R}{r}\right)\left(\Gamma \chi\left(\theta \right)  {\cal I}-1\right)    \nonumber \\ 
&+& \frac{1}{2}{\Omega}^{2}\sin^{2}\theta\left[1-\left(\frac{R}{r}\right)^{2}\right]
\label{Vr}
\end{eqnarray}
where
\begin{eqnarray}
 {\cal I} &\equiv& \frac{1-e^{-m  \chi\left(\theta \right) \left(1-{R}/{r}\right)}}{m\chi\left(\theta \right)\left(1-{R}/{r}\right)}       \\
  &=& 1 - \frac{m\chi(\theta)}{2} \left(1-\frac{R}{r}\right)+\mathcal{O}\left(m^{2}\right).\nonumber
\end{eqnarray}
Note that the wind approaches its terminal velocity after traversing just a few stellar radii:
\begin{equation}
\left(\frac{v_{r}}{v_{\infty}}\right)^{2} > 1-\frac{R}{r}.
\label{eq:TerminalVelocity}
\end{equation}
This justifies our second assumption, that $v_{\theta}\ll v_r$. In principal, one could argue that even if the star were spherical, radiative diffusion would have given rise to latitudinal radiation and velocity components. However, since most of the acceleration takes place close to the surface, these components are bound to be small.
 
\subsection{Self Spin-up by Continuum-driven Winds}
\label{sec:Spinup}

As the star blows wind, it loses angular momentum. However, specific angular momentum and thus the dimensionless spin $\Omega$ may increase, as we now show. 

We start with conservation of angular momentum. The angular momentum reduction of the star is the angular momentum taken by the wind:
\begin{equation}
\label{eq:lconserve}
\dot{\ell}_s =-\dot{\ell}_w,
\end{equation}
where
\begin{equation}
\ell_s = \omega M R^2 \alpha_g^2,
\end{equation}
and 
\begin{eqnarray}
\dot{\ell}_w&=&\omega \int R^2\sin^2\theta \,{\bf \Phi}\cdot {\bf dS}  =-\omega \dot M  R^2 \frac{1}{2}\int\chi\sin^3\theta d\theta  \nonumber \\ 
&\equiv &-\omega \dot M  R^2 \alpha^2_w. 
\end{eqnarray}
Here $\alpha_g$ is the stellar radius of gyration. $\alpha_w$ is an effective radius of gyration of the wind, defined in the equation above. 

Dividing eq.~(\ref{eq:lconserve}) by $\ell_s$,  
we find
\begin{equation}
\label{eq:omegaevol}
\frac{3}{2} \frac{\dot M}{M}+\frac{1}{2} \frac{\dot R}{R}+\frac{\dot \Omega}{\Omega}=\frac{\alpha_w^2}{\alpha_g^2} \frac{\dot{M}}{M}. 
\end{equation}
In the upper part of the main sequence, where radiation pressure dominates, one roughly has that $R \propto \sqrt{M}$, such that $2 {{\dot R}/{R}} \approx {\dot M}/{M}$. Eq.~(\ref{eq:omegaevol}) can now be integrated to give
\begin{equation}
\log\frac{\Omega_f}{\Omega_i}=\left( \frac{\alpha_w^2}{\alpha_g^2}-\frac{7}{4} \right) \log{\frac{M_f}{M_i}}.
\label{eq:loglog}
\end{equation}
This assumes that over the integration interval, $\alpha_w$ remains constant.\footnote{This assumption is applicable to both
 the present day wind with its low mass loss rate, and high load winds at a steady state. Eruptions are treated differently in the following section.}
The condition for self spin-up is therefore 
\begin{equation}
\alpha_w^2 < \frac{7}{4} \alpha_g^2.
\label{eq:selfspinup}
\end{equation}

For a high load wind $\alpha^2_w\approx 2/5$, 
and the star spins down.\footnote{Massive stars typically have
$\alpha_g^2\sim0.1$ \citep{Motz1952}. } However, at low mass loss rates, corresponding to $m\ll 1$, the flux is insufficient to push the wind  over the effective potential at equatorial latitudes. 
The wind reaches infinity only for angles between the pole and  
\begin{equation}
\theta_{\textrm{max}}\approx\sqrt{m\frac{2/{\tilde{\mathcal{W}}}-1}{\Omega^{2}\left(1-\frac{2}{3}\Omega^{2}\right)}} +\mathcal{O}(m^{3/2}).
\label{eq:thetamax}
\end{equation}
We expect the rest of the wind to stagnate and fall back, thus taking no net angular momentum. 
Therefore,
\begin{eqnarray}
\alpha^2_w&=&\int_0^{\thetamax}\chi\sin^3\theta d\theta \\ &=&\frac{1}{4}\left( 1-\frac{2}{3}\Omega^2\right)^{-1}\thetamax^4+\mathcal{O}(\thetamax^6).  \nonumber 
\label{eq:alphaw}
\end{eqnarray}
One should note that the photon tiring number should be computed using the mass loss at the surface of the star, not at infinity. This quantity relates to the observed ejecta by truncating the surface integration of $\bf \Phi$ (eq. \ref{eq:MdotPhi}) at $\theta_{\textrm{max}}$:
\begin{eqnarray}
m_\infty&=&m\int _{0} ^{\theta_{\textrm{max}}} \chi \sin \theta d \theta \\&=&m^2\frac{2/\tilde{\mathcal{W}}-1}{2\Omega^2\left(1-\frac{2}{3}\Omega^2 \right)^2}+\mathcal{O}\left(m^3\right). \nonumber 
\label{eq:minfty}
\end{eqnarray}

Using equations (\ref{eq:thetamax}) - (\ref{eq:minfty}) one finds
\begin{equation}
\alpha_w^2\approx m_\infty \frac{ 2/\tilde{\mathcal{W}}-1}{2\Omega^2\left(1-\frac{2}{3}\Omega^2 \right)}.
\label{eq:alphaVSm}
\end{equation}

\subsection{Spin Evolution During Eruptions}
\label{sec:SkirtTheory}

In the previous section, we studied the effects of continuum driven
winds on the evolution of angular momentum assuming a constant luminosity.
As we shall now see, an abrupt change in luminosity has significant
consequences regarding the evolution of the spin of the star. 

A sudden increase in luminosity can be understood as a result of an
atmospheric phase transition, where the atmosphere becomes porous 
and the effective opacity drops \citep{ShavivNovae}. Consequently, the radiative flux at
the surface becomes greater than the incoming convective flux.
In order to compensate for the imbalance, the star must contract while
radiating the difference in  binding energy $U$:\footnote{Because of its low density, the atmosphere alone does not have sufficient
gravitational binding energy. Note that the increase in luminosity is associated only with the release of binding energy due to contraction, which will eventually stop once nuclear reactions set in.}
\begin{equation}
\Delta L=L_{\textrm{erup}}-L_{\textrm{init}}=-\dot{U}\big|_M=-\frac{\partial U}{\partial R} \dot{R},
\end{equation}
where $L_{\textrm{erup}}$ and $L_{\textrm{init}}$ are the luminosities
during and before the eruption, respectively. A more convenient choice of parameters is $L_{\textrm{erup}}$ and $\lambda\equiv 1-L_{\textrm{init}}/L_{\textrm{erup}}$; 
henceforth, $L_{\textrm{erup}}$ will be denoted by $L$.

In order to keep this discussion as generic as possible, we parametrize
the binding energy as
\begin{equation}
U\equiv-\frac{GM^{2}}{R}\mathcal{B}
\end{equation}
and assume a constant $\mathcal{B}$, which is generally the case.  The time scale
for contraction is then
\begin{equation}
-\frac{R}{\dot{R}}=T_{\textrm{KH}}\mathcal{B}\lambda^{-1},
\end{equation}
where $T_{\textrm{KH}}\equiv G M^2 /L R$ is the Kelvin-Helmholtz time scale. Plugging
this into eq.~(\ref{eq:omegaevol}) gives 
\begin{equation}
\frac{\lambda}{2}\left(T_{\textrm{KH}}\mathcal{B}\right)^{-1}+\left(\frac{\alpha_{w}^{2}}{\alpha_{g}^{2}}-\frac{3}{2}\right)\frac{\dot{M}}{M}=\frac{\dot{\Omega}}{\Omega}.
\label{eq:AngMomCont}
\end{equation}
The contraction terminates when the  temperature build up at the
core becomes sufficient to generate the required luminosity through nuclear reactions. 

In order to estimate how long does the star need to contract, we assume that it behaves homologously and consider 
 the scaling of the luminosity. If the specific energy production can be written as $\epsilon \propto \rho^p T^q$, we get
\begin{equation}
  L = \int  \epsilon \rho\, dV  \propto \rho_c^{1+p} T_c^{q} R^3.
\end{equation}

In the following, we assume that $p=1$, which is suitable for most nuclear reactions, Hydrogen burning included. 
The temperature scales as $T \propto P/\rho$ when gas pressure dominates, and $T\propto P^{1/4}$ when radiation pressure dominates.
 If the dynamical time scale, $T_{\textrm{dyn}}\equiv 1/\sqrt{G\rho}$, is much shorter than the contraction time, one can safely assume mechanical equilibrium.\footnote{ For the case of \etacar, $T_{\textrm{dyn}}\approx$ 2 weeks.} Hydrostatics then give that $P\propto M^2/R^4$. 
We thus find $L \propto M^{2+q/\nu} R^{-3-q}$ with $\nu=1 ( \mathrm{or}~ 2)$, corresponding to the case where gas (or radiation) pressure dominates.

Because $q$ is typically very large, the star needs to contract only by a relatively small amount
\begin{equation}
\frac{\Delta R_{\textrm{cont}}}{R}  = 1 - \left( L_{\textrm{init}} \over L_{\textrm{erup}} \right)^{1/(3+q)}\left(1-\frac{\Delta M}{M}\right)^{(2+q/\nu)/(3+q)}.
\label{eq:ContractionLength}
\end{equation}
The contraction will therefore take place for a duration
\begin{equation}
T_{\textrm{cont}}=\int \frac{dR}{\dot R}\approx T_{\textrm{KH}}\mathcal{B}\lambda^{-1}\frac{\Delta R_{\textrm{cont}}}{R}.
\label{eq:ContractionTime}
\end{equation}

\section{Part II - Application to $\eta$-Carinae}
\label{Part2}

We now show how the theory developed in \S\ref{Part1} applies to the particular case of \etacar.  In \S\ref{sec:WindApplication}, it is used to estimate the shape of the SED outflow ejected during the Great Eruption. We argue that the Homunculus' shape is a natural consequence of a star rotating near breakup with SED continuum driven winds. In \S\ref{sec:presentwind}, we show that the present day wind is sufficiently concentrated towards the poles to spin up the star towards critical rotation, if it is the dominant mode of mass loss. 
 In \S\ref{sec:SkirtApp}, we show that also the formation of an equatorial skirt is expected given the stellar parameters of \etacar. 

We consider nominal values of $M=120 M_{\odot}$, $R=150R_{\odot}$ for the mass and radius. 
Since the age of the star is unknown, we use an intermediate chemical composition corresponding to $\kappa= 0.3\ \textrm{cm}^{2}\textrm{g}^{-1}$ for the opacity.  As advocated, the star is assumed to be near critical rotation.

\subsection{The Homunculus as a Continuum-driven Wind from a Rotating Star}
\label{sec:WindApplication}

We assume that during the Great Eruption, the sustained mass loss rate had been of the order of $\dot{M}=-0.1 M_{\odot}\textrm{year}^{-1}$.\footnote{More about this choice will be discussed below.} Given the aforementioned nominal stellar parameters, the corresponding photon tiring number is $m\approx0.5$. Assuming critical rotation, the size of the Homunculus is reproduced by a value of $\Gamma\approx1.6$, which in turn implies a scaled wind constant of $\tilde{\mathcal{W}}\approx0.4$.  
This value of $\Gamma$ corresponds to a luminosity~$L=\Gamma L_{\textrm{edd}}\approx 10^{7} L_{\odot}$ at the base of the wind. 

At infinity, the modeled luminosity must account for photon tiring:
\begin{equation}
L_{\infty}=\intop {\bf F}\cdot {\bf dS}= L\intop  \frac{d\Omega}{4\pi}e^{-m \chi(\theta)}\chi(\theta)\approx 0.5 L.
\end{equation}
For comparison, the observed luminosity at the time was in the range of $L_{\textrm{obs}}\approx 1-3 \times 10^{7} L_{\odot}$ \citep{SmithFrew}.

\begin{figure}[t]
\centerline{
\epsfig{file=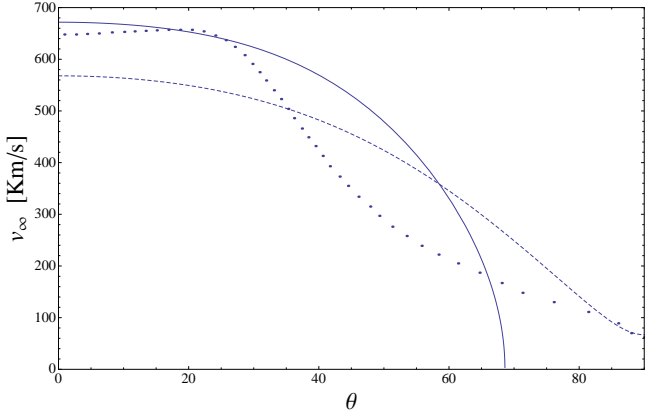, angle=0,width=8.5 cm}
}
\caption{
\footnotesize{The latitudinal dependence of the radial velocity of the wind. The dots correspond to the values observed by   \cite{Smith2006}. The solid line represents the model described in \S\ref{sec:WindsTheory} for the nominal stellar parameters and $\Gamma\approx1.6$ at critical rotation. The dashed line corresponds to the same model with $\Omega\approx0.9$ -- the velocity profile expected after a few years of mass loss at a rate of $\dot{M}=-0.1 M_{\odot}\textrm{year}^{-1}$. 
$\qquad\qquad\qquad\qquad\qquad\qquad\qquad\qquad\qquad\qquad\qquad\qquad$
}
}
\label{VofTheta}
\end{figure}
Figure \ref{VofTheta} shows the latitudinal dependence of the wind velocity predicted by eq.~(\ref{Vr}) versus observed values.

\subsection{Spin-up from the Present-day Wind}
\label{sec:presentwind}

\begin{figure}[t]
\centerline{
\epsfig{file=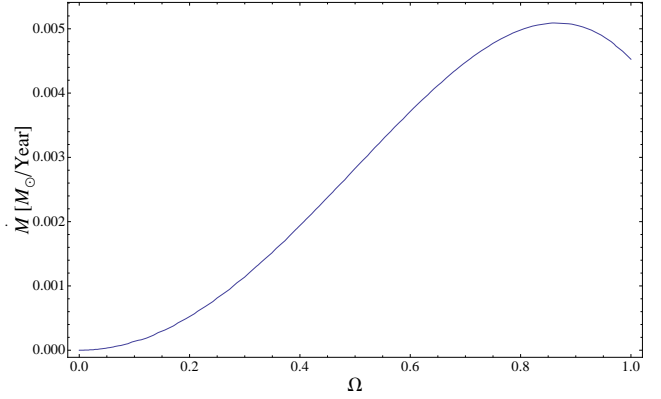, angle=0,width=8.5cm}
}
\caption{
\footnotesize{The spin-up condition (eq. \ref{eq:selfspinup}) is satisfied for values of spin and mass loss rate below the line.}
}
\label{fig:spinup}
\end{figure}

The present day mass loss rate is of the order of $10^{-3} M_{\odot}\textrm{year}^{-1}$ \citep{DavidsonReview}, which implies 
$\alpha_w^2\approx3\times10^{-2}$ (eq. \ref{eq:minfty}). 
In this mode of mass loss, it would take \etacar\ about $2\times10^3$ years to reach again critical rotation, for the above nominal parameters. 
The maximal mass loss rates able to satisfy eq.~(\ref{eq:selfspinup}) as a function of spin can be read off Figure \ref{fig:spinup}.

\subsection{The Formation of the Equatorial Skirt}
\label{sec:SkirtApp}

Any attempt to model the dynamics behind the Homunculus nebula cannot
be complete without  explaining the formation of the equatorial skirt.
As it turns out, the formation of such a skirt is a generic property of SED eruptions
at critical rotation.

Consider the equation of conservation of angular momentum for the
contracting star (eq.~\ref{eq:AngMomCont}). If the star erupts while critically rotating and 
\begin{equation}
\frac{M}{T_{\textrm{KH}}}>|\dot{M}|\left[2\lambda
^{-1}\mathcal{B}\left(\frac{\alpha_{w}^{2}}{\alpha_{g}^{2}}-\frac{3}{2}\right)\right],
\end{equation}
the equator becomes unbound,
 and a skirt is formed at a rate:\footnote{This can be seen by replacing $\dot{\Omega}/\Omega$ by $\dot{M}_{\textrm{sk}}/M\alpha_{g}^{2}$
in eq. (\ref{eq:AngMomCont}). 
} 
\begin{equation}
\dot{M}_{\textrm{sk}}=\alpha_{g}^{2}\left[\frac{\lambda}{2}\mathcal{B}^{-1}\frac{M}{T_{\textrm{KH}}}+\left(\frac{\alpha_{w}^{2}}{\alpha_{g}^{2}}-\frac{3}{2}\right)\dot{M}\right].
\label{eq:MdotSkirt}
\end{equation}
Note that the mass loss to the skirt is neglected with respect to the mass loss to the wind; this is justified by self consistency.

\begin{figure}[t]
\centerline{
\epsfig{file=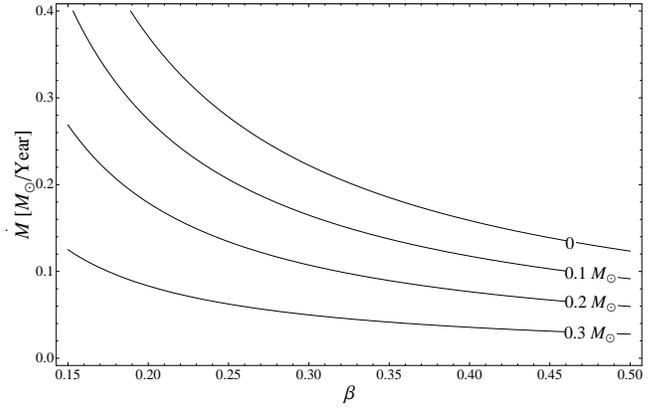, angle=0,width=8.5cm}
}
\caption{
\footnotesize{The mass of the equatorial skirt, as a function of 
$\dot{M}$ and $\beta\equiv P_{\textrm{gas}}/P_{\textrm{tot}}$. 
The values given here should be considered as lower limits, as the relevant radius of gyration of the outer shell participating in the dynamics is much larger than that of the whole star, see \S\ref{sec:AngMomDif}. 
}
}
\label{fig:SkirtMass}
\end{figure}

For a critically rotating $n=3$ polytrope, which is a reasonable description of \etacar, one has that
\begin{equation}
\mathcal{B}=\frac{3}{4}\beta-\frac{1}{2}\alpha_{g}^{2},
\end{equation}
where $\beta\equiv P_{\textrm{gas}}/P_{\textrm{tot}}$  \citep[see, for example,][]{Clayton1968}.

\cite{ShavivInstabilities} has shown that the radiative hydrodynamic instabilities, which are responsible for the atmospheric phase transitions, are excited for $0.15 \lesssim \beta \lesssim 0.5$, depending on different conditions. In this range of $\beta$, one has $\nu \approx 2$.

Comparing the luminosities computed in the previous two sections, we see that during the eruption, the luminosity increased roughly by a factor of $5$, and so $\lambda\approx4/5$.

The last piece of information required to understand the contraction is the factor $q$, relating the specific energy production and temperature by  $\epsilon \propto T^q$. Since the temperature at the core of \etacar\ is expected to be of the order of $5\times 10^7$K, the energy production is due to CNO burning, which at this temperature corresponds to $q \sim 15$ \citep[e.g.,][]{Bethe1939}. 

We can now estimate (eq.~\ref{eq:ContractionTime}) the contraction time to be $3$ to $10$ years, depending on the value of $\beta$, and so the skirt's mass is of the order of $0.1\, M_{\odot}$. Figure \ref{fig:SkirtMass} shows the predicted mass of the skirt for different values of $\beta$ and $\dot{M}$. 

Throughout this analysis, we have implicitly assumed that the whole star
is instantly affected by the loss of angular momentum to the wind, and thus
the corresponding $\alpha_{g}^{2}$ is constant. However, during time
periods shorter than the time scale of angular momentum diffusion,
only an outer shell plays a role in the dynamics. In \S\ref{sec:AngMomDif}, we show that the diffusion time scale is indeed comparable if not longer than the contraction time.  As a result,
the effective radius of gyration is larger, and so the mass
of the skirt computed from eq. (\ref{eq:MdotSkirt}) is in fact a lower limit.
We conclude that during the initial stages of contraction, a skirt is likely to have formed even for relatively high mass loss rates.
\newpage

\section{Discussion and Summary}
\label{sec:Conclusions}

The twin lobe structure of the Homunculus nebula of \etacar\ strongly suggests that either fast rotation or binary interaction played an important role in the process of mass loss. In this work, we have chosen to pursue the idea that it is near critical rotation that sculpted the Homunculus, and that the mass loss was in the form of a continuum driven wind. The formation of the equatorial skirt is then associated with the evolution of angular momentum before the system settled in a steady state.

As we have seen in \S\ref{sec:WindsTheory}, the structure of such winds can be obtained analytically, if one allows several simplifying assumptions. In particular, it was assumed that there is no latitudinal interaction between mass and radiative flux elements on different radial trajectories. This assumption later proved to be self consistent, as it turns out that most of the acceleration of the wind takes place near the surface.\footnote{Eq.~(\ref{Vr}) implies that the winds of \etacar\ reach half their terminal velocity already at $r \approx 1.15 R$. For comparison, line driven winds reach half their terminal velocity further out, at $r \approx 1.73R$ \citep[e.g.,][]{Lamers1999}.} Namely, the relevant radial scale is smaller than the latitudinal one; this acts to suppress lateral fluxes.  

By further simplifying the geometry and assuming a spherical star, the set of equations describing all conserved quantities became both algebraically closed and integrable. Obviously, this assumption cannot hold since rotation breaks spherical symmetry. Even so, we find that this simple model captures the gross features of the system, as depicted in figure~\ref{VofTheta}. Mainly, a twin-lobed structure emerges as a generic property of winds from such stars. The main ingredient required in order to form this structure is that of gravity darkening, as was already pointed out by \cite{Owocki1998} and \cite{Maeder1999} for line driven winds. Nevertheless, we do expect the oblate geometry to introduce distortion. For example, the predicted velocity may be lower near the equator, because the wind is launched from larger radii. This may in fact improve the agreement with the observations.

For the specific case of \etacar\ (\S\ref{sec:WindApplication}), our model provides a rough description of the shape and size of the Homunculus nebula.
In particular, continuum driven winds naturally explain the large wind momentum to photon luminosity observed in the Homunculus. Such high ratios cannot be explained by line driven winds. 

Note however, that within the approximations and the nominal parameters used, the typical mass of the Homunculus obtained is somewhat lower than the value recent observations imply \citep{Smith2003}, and the predicted luminosity is lower by a factor of $\sim 3$.  
These discrepancies may be reduced by a full numerical analysis.

As a side note, we show in \S\ref{sec:spin} that the angular momentum gain due to a general spin-orbit coupling is correlated with the change in semi-major axis. As a consequence, in light of the high eccentricity of the orbit and stellar parameters assumed, the angular momentum \etacar\ may have received from interactions with its companion were insufficient to drive it to critical rotation. 

The fact that the rotation is very close to critical may seem like fine tuning, but it is quite possible that this is the natural state of the system. Since the star cannot sustain high load winds for long periods of time, it is reasonable that a low mass loss rate, as \etacar\ currently exhibits, is in fact the dominant mode of mass loss. Motivated by this reasoning, we analyzed in \S\ref{sec:Spinup} the explicit solutions for the
wind velocity at low mass loss rates. We found that at equatorial latitudes, the wind stagnates and falls back to the surface, given that the star is rotating fast enough. 
The mass loss from pole centered winds, which do reach infinity, dominates over the loss of angular momentum and so the star spins up.\footnote{By ``spin-up" we refer to an increase of 
$\Omega\equiv\omega\,\sqrt{R^3/GM}$.} 
This is another aspect of continuum driven winds which is absent from the line driven winds model. 

We predict (\S\ref{sec:presentwind}) that in about $2\times10^3$ years at the current mass loss rate, \etacar\ will be back at critical rotation.  This implies that if the previous large eruption cycle and subsequent wind were similar to the Great Eruption and the present wind, then the previous large eruption of \etacar\ must have taken place at least a few thousand years ago. 

We then showed in \S\ref{sec:SkirtTheory} that at the onset of LBV eruptions, the contraction that follows the atmospheric phase transition is fast enough to render equatorial latitudes unbound, if the star is critically rotating. The formation of an equatorial skirt is seen to be a generic consequence of the model as well. For the stellar parameters of \etacar\ (\S\ref{sec:SkirtApp}), we predict the skirt's mass to be of the order of $0.1\,M_\odot$. 

The effect of angular momentum diffusion, which is discussed in \S\ref{sec:AngMomDif}, complicates the estimation of the mass of the skirt. Taking it into account requires a more accurate knowledge of the mass loss to the wind, and specifically its time dependence. Essentially, the question is what happened first - did nuclear reactions kick in to stop the contraction or did the loss of angular momentum manage to diffuse inwards and render the formation of a skirt unnecessary? We await the appearance of new observational data regarding the time dependence of the Great Eruption. 

\section*{Acknowledgments}
TS is supported by the BSF \textendash{} American-Israel Bi-National Science
Foundation, and by a center of excellence supported by the Israel
Science Foundation,~grant~1665/10. 
NJS is supported by the Israel Science Foundation, grant 1589/10.

\appendix

\section{A Bound on the Angular Momentum Gained from Spin-Orbit Coupling}
\label{sec:spin}

The above analysis assumed that there is no additional exchange of angular momentum. However, \etacar\ is part of a binary system, and one might suspect that spin-orbit coupling plays a dominant role in its spin evolution. Given the orbital parameters, we argue that such interactions do not offer a significant contribution to the spin, and can therefore be neglected.

The basic idea is that the change in the spin is correlated with a change in the semi-major axis; any angular momentum conserving interaction 
cannot change one without effecting the other. We show that even if the original binary orbit had an extremely large semi-major axis, the evolution to the present orbit could not have spun up \etacar\ significantly.

Let us then consider a general force {\bf f} representing the interaction. Assuming the interaction is important only at short distances,\footnote{For example, tidal interactions give that ${\bf f}\propto r^{-7}$ \citep{Hut1981}.}
 it may be approximated as an impulse during periastron passages. 
 $\\$Due to conservation of total angular momentum, the comparison between the spin angular momentum, $\ell_{s},$ and the orbital angular momentum, $\ell_o$, gives
\begin{equation}
\dot{\ell}_{s}=- \dot{\ell}_{o}=- \left(f_{\theta}r+\dot{r}r\mu\dot{\theta}\right),
\label{eq:ells}
\end{equation}
where $r(\theta)$ describes the orbital motion. We further neglect the term $\dot{r}r\mu\dot{\theta}$ since $\dot{r}\rightarrow0$ at periastron. In order to reduce the clutter of notation, $\ell$ will be henceforth used to denote $\ell_{s}$.  

During one orbit, the change in the 
angular momentum is 
\begin{equation}
\delta\ell \approx\intop_{p} \dot{\ell}dt
\approx\dot{\ell}_{p}\intop_{p}dt. 
\end{equation}
where the subscript $p$ denotes values calculated at periastron.\footnote{We use $\delta$ to denote changes over one orbit, and we will use $\Delta$ for changes over the whole evolution.} The change in the semi-major axis is due to the loss of orbital energy, $E=-GMm/2a$, hence
\begin{equation}
\delta a = \frac{2a^{2}}{GMm} \delta E.
\end{equation}
 On the other hand, the change in energy is
\begin{equation}
\delta E=\int {\bf f }\cdot{\bf dr} 
\approx \intop_{p}  f_{\theta} rd\theta
\approx -\dot{\ell}_p \dot{\theta}_{p} \intop_{p} dt.
\end{equation}
One should note that the periastron separation, $r_{p}=a(1-\epsilon)$, is approximately constant during orbital evolution at high eccentricities. Since the interaction takes place at periastron, the evolution of the orbit is such that the periastron remains fixed while the apastron, $r_a$, progressively shrinks. Although the orbit may change during periastron passages when the force is applied, the periastron itself must remain part of it.  
This implies that the angular velocity at periastron passages,
 \begin{equation}
 \dot{\theta}_{p}=\sqrt{r_{p}^{-3}G(M+m)(1+\epsilon)},
\end{equation}
changes very slowly and 
is assumed to be constant as well.

On a long time scale, the ratio between time derivatives of the semi-major axis and the spin is the ratio between the corresponding incremental changes every periastron passage:
\begin{equation}
 \frac{\dot{a}}{\dot{\ell}}\approx \frac{\delta a}{\delta \ell}=-\frac{2a^{2}\dot{\theta}_{p}}{GMm}.
\end{equation}
This expression can be integrated to give
\begin{equation}
\label{eq:deltaell}
\Delta \ell \equiv \ell\Big|_{0}^{t}=\frac{GMm}{2 \dot{\theta_{p}} }a^{-1}\Big|_{0}^{t}.
\end{equation}
Clearly, the maximum possible change in the angular momentum is 
\begin{equation}
\label{eq:deltaellmax}
\Delta \ell_{\textrm{max}} = { GMm \over 2\dot{\theta_{p}}a}.
\end{equation}
 Namely, the upper bound depends only on the current value of the semi-major axis.
	
Another way of looking at the orbital evolution is global.  We can compare the angular momentum of two systems having the same periastron, $r_p$, but one having $r_a \rightarrow \infty$, and the other an $r_a$ of the present orbit. In contrast to the previous derivation, a global view cannot explain why the perisatron distance remains approximately constant during the evolution. 

The general expression for the angular momentum of a Keplerian orbit is
\begin{equation}
\ell_o^{2}=G\frac{M^{2}m^{2}}{M+m}r_{p}(1+\epsilon).
\end{equation}
We define $\Delta \ell_{o,\textrm{max}}\equiv \ell(\epsilon)-\ell(\epsilon\rightarrow1)$ and expand the angular momentum for small values of $1-\epsilon$. 
We then obtain, to the leading order,
\begin{equation}
\left| \Delta \ell_{o,\textrm{max}} \right|  \leq \frac{GMm}{2\dot{\theta}_{p}a},
\end{equation}
in agreement with eq.~(\ref{eq:deltaellmax}).

This change of angular momentum should be compared with the present value of the angular momentum of the star, while assuming it is rotating near breakup.
For the present nominal stellar parameters, including a period of $5.5$~years and $\epsilon=0.9$, we find that 
\begin{equation}{\Delta\ell_{\textrm{max}}}/{\ell_{s}}\sim 5\%,\end{equation} and thus interaction with the companion star could not be held responsible for the high spin of \etacar. 

\section{Angular Momentum Diffusion}
\label{sec:AngMomDif}
\def\dphidxi1{\left. d\phi / d\xi\right|_{\xi_1} / \xi_1 }
\def\dphidx{\left. d\phi / d\xi\right|_{\xi_1}  }
The last theoretical calculation required for the understanding of angular momentum evolution in mass losing stars is an analysis of angular momentum diffusion. Until now, we have implicitly assumed that the whole star immediately reacts to the mass loss. However, the reaction time cannot be shorter than the diffusion time scale $T_\textrm{diff}$.  

When considering processes
lasting for time scales comparable to $T_\textrm{diff}$, the star cannot be thought of as a rigidly rotating body. 
Since only an outer shell  participates in the dynamics, the effective radius of gyration is much larger than that of the entire star. 

The goal of this section is to obtain a lower bound for $T_\textrm{diff}$, the diffusion time scale associated with the homogenization of such a shell. As we shall see, the angular momentum diffusion time in massive stars ranges from a fraction of year for the outer parts of the envelope, to several years to homogenize the whole star.

Since $T_\textrm{diff} \sim{R^2}/{ {\mathcal{D} }}$, we seek an upper bound on the diffusion coefficient $\mathcal{D}$. We note that the most efficient diffusion mechanism in stars is that of convective mixing \citep{Taylor1915}.\footnote{Other mechanisms such as shear induced turbulence or meridional circulation are important when convection is absent \cite[e.g.,][]{Zahn1992}.} We also note that convection is always excited above some critical, sub-Eddington luminosity \citep{JSO}, implying that stars with continuum driven winds are necessarily convective. This motivates the application of mixing length theory  \citep[MLT,][]{Vitense1953,Bohm1958} to angular momentum transport in a SED state. 

Under MLT, the diffusion coefficient is
\begin{equation}
\mathcal{D} \approx {h v_c \over 3},
\end{equation}
where $v_c$ is the convective velocity and $h$ is the pressure scale height, playing the role of the mean free path. 

MLT also gives that the convective velocity is 
\begin{equation}
v_c \approx  \left( L \over 4 \pi r^2 \alpha \rho  \right)^{1/3},
\label{eq:vc}
\end{equation}
where  $\alpha = c_p \rho / v_s^2 \approx 3/2$ and $c_p$ is the heat capacity per unit mass. 
Since $v_c$ is obviously limited by the speed of sound, eq.~(\ref{eq:vc}) should be regarded as an upper bound. 

\begin{figure}[t]
\centerline{
\epsfig{file=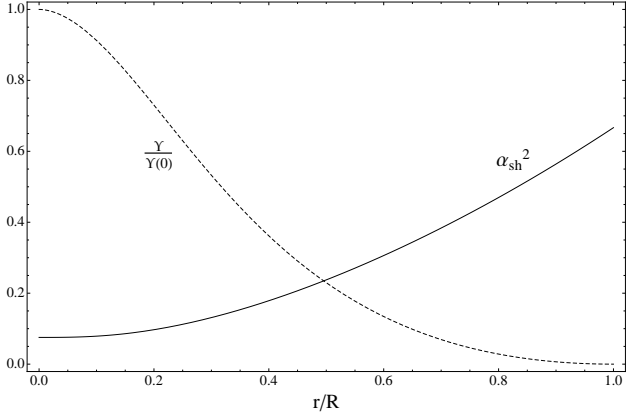, angle=0,width=8.5cm}
}
\caption{
\footnotesize{ 
The solid line depicts the squared radius of gyration of a shell located between a certain distance from the center, $r$, and the stellar surface, $R$. The dashed line plots the dimensionless function $\Upsilon$ describing the typical time it takes to convectively diffuse angular momentum throughout the shell. These functions are plotted for an $n=3$ polytrope.  }
}
\label{fig:upsilon}
\end{figure}

In order to compute the scale height, we model the star as an $n=3$ polytrope\footnote{For completeness, we briefly review the polytropic approximation: the pressure and density are conjectured to be related by $\rho=\rho_c \phi^n$ and $P=P_c \phi^{n+1}$ for some value of $n$. The hydrostatic equation, $\nabla P=-\rho\nabla\Phi$, then reads $(n+1)P_c\nabla\phi=-\rho_c\nabla\Phi$, and Poisson's equation for the gravitational potential, $\nabla^{2}\Phi=4\pi G\rho$, becomes the Lane-Emden equation \citep{Lane1870,Emden1907}:
\begin{equation}
\frac{1}{\xi^{2}}\frac{d}{d\xi}\left(\xi^{2}\frac{d}{d\xi}\phi\right)=-\phi^{n},
\end{equation}
where spherical symmetry is assumed and $\xi \equiv r \sqrt{4\pi G \rho_c^2(n+1)^{-1}P_c^{-1}} $.
$\xi_1$ is the value for which $\phi= 0$; for $n=3$, a numerical computation gives $\xi_1\approx 6.9$ . Note that $\xi_1$ is related to the physical radius through $r = (\xi/\xi_1) R$.
}
 \citep{Eddington1931}:
 \begin{equation}
h \equiv \left( d \ln P \over d r \right)^{-1} = R\left[ {\phi \over n+1}\frac{1}{ \xi_1}\left( {d \phi \over d\xi}\right)^{-1}\right].
\end{equation}
The diffusion coefficient is then
\begin{equation}
\mathcal{D} = \left( L\, R^{4} \over M\right)^{1/3}\left[ {  \left(-\dphidx\right)^{1/3} \over 3 (1+n) \alpha^{1/3} \xi_1^{2/3}  }  { \phi^{1-{n\over3}} \over{ \xi^{2/3}} } \left( {d \phi \over d\xi}\right)^{-1} \right]
\end{equation} 
and the typical diffusion time between a radius $r$ and the surface of the star is approximately given by
\begin{equation}
T_{\textrm{diff}} \approx\left(\int_r^R {dr \over \sqrt{\mathcal{D} }} \right)^2=\left(\frac{ M\, R^{2}}{ L }\right)^{1/3} \Upsilon(\xi),
\end{equation}
where we have defined 
\begin{equation}
\Upsilon(\xi) \equiv \left( \int_\xi^{\xi_1} \phi^{(n-3)/6} \sqrt{ \frac{d\phi}{  d\xi}} \xi^{1/3} d\xi\ \right)^2 \left(\frac{ 27 (1+n)^3 \alpha}{  -\xi_1^4\dphidx }\right)^{1/3}.
\end{equation}
For the stellar parameters of \etacar, the diffusion time is
\begin{eqnarray}
T_\textrm{diff} &\approx&  10\ \mathrm{years} \times\\ &&\left( M \over 120 M_\odot \right)^{1/3} \left( R \over 150 R_\odot \right)^{2/3} \left( L \over 10^7 L_\odot \right)^{-1/3}  \frac{\Upsilon}{\Upsilon(0)}. \nonumber
\end{eqnarray}
$\\$
The radius of gyration of the outer shell is given by 
\begin{equation}
\alpha_{\textrm{sh}}^2(\xi) = { \int_{\textrm{sh}} r_{\perp}^2 dM  \over \int_{\textrm{sh}} dM}  = {2 \over 3}  { \int_\xi^{\xi_1} \xi^4 \phi^n d\xi \over  \xi_1^2 \int_\xi^{\xi_1} \xi^2 \phi^n d\xi }.
\end{equation}
The functions $\Upsilon$ and $\alpha_{\textrm{sh}}^2$
 are plotted in figure \ref{fig:upsilon}. 

$\\$
$\\$
$\\$


\end{document}